# Noise suppression in a temporal-multimode quantum memory entangled with a photon via asymmetrical photon-collection channel

Ya Li[1,2], Ya-fei Wen[1,3], Min-jie Wang[1,2], Chao Liu[1,2], Hai-long Liu[1,2],

Shu-jing Li[1,2], Zhong-xiao Xu[1,2], Hai Wang[* 1,2]

[1]The State Key Laboratory of Quantum Optics and Quantum Optics Devices, Institute of Opto-Electronics, Shanxi University, Taiyuan 030006, China

[2]Collaborative Innovation Center of Extreme Optics, Shanxi University, Taiyuan 030006, China

[3]Department of Physics, Taiyuan Normal University, Jinzhong 030619,China

*Corresponding author: wanghai@sxu.edu.cn

## ABSTRACT

Quantum interfaces (QIs) that generate entanglement between a multimode atomic memory and a photon forms a multiplexed repeater node and hold promise to greatly improve quantum repeater rates. Recently, the temporal multimode spin-wave memory that is entangled with a photon has been demonstrated with cold atoms. However, due to additional noise generated in multimode operation, the fidelity of spin-wave-photon entanglement significantly decreases with the mode number. So far, the improvement on temporal-multimode entanglement fidelity via suppressing the additional noise remains unexplored. Here, we propose and experimentally demonstrate a scheme that can suppress the additional noise of a temporally-multiplexed QI. The scheme uses an asymmetric channel to collect the photons coming and retrieving from the temporally-multiplexed QI. For making comparisons, we also set up a QI that uses symmetric channel for the photon collections. When the QIs store 14 modes, the measured Bell parameter $S$ for the QIs using the asymmetric and the symmetric photon-collection channels are 2.36±0.03 and 2.24±0.04, respectively, showing that the QI using the asymmetric channel gives rise to a 3% increase in entanglement fidelity, i.e., a 1.7-fold decrease in the additional noise, compared with the QI using the symmetric one. On the other hand, the 14-mode entanglement QIs that use the asymmetric and symmetric collections preserve the violation of a Bell inequality for storage times up to 25 μs and 20 μs, respectively, showing that the asymmetric QI has a higher entanglement storage performance.

## INTRODUCTION

The large-scale quantum networks[1-3] and long-distance quantum communication[3-5] rely on long-distance entanglement distribution through quantum repeaters (QRs) [6]. In QRs[4-6], a long distance $L$, over which one wants to distribute entanglement, is divided into short elementary links. Each link contains two nodes with a separated distance $L_0$. Entanglement is generated on each short interval, and then extended to the whole distance $L$ through entanglement swapping[6]. The key element of each node is a light-matter quantum interface (LMQI) that generates quantum correlations or entanglement between an atomic memory and a photon[4-5]. To practically realize QRs, Duan-Lukin-Cirac-Zoller (DLCZ) proposed a protocol[5], where, atom-photon quantum correlations can be created via spontaneous Raman scattering (SRS) induced by a write laser pulse in atomic ensembles. The atom-photon quantum correlations have been demonstrated with atomic ensembles[7-20]. Building on the atom-photon quantum correlations, the generations of atom-photon entanglement have been demonstrated in many experiments[21-31]. Using atom-photon entanglement instead of atom-photon quantum correlation as the repeater nodes, the long-distance phase stability, required in DLCZ protocol[5] is no longer necessary. Alternatively, the atom–photon entanglement quantum interface (QI) may be generated through the storages of photonic entanglement[32-34] in atomic ensembles. With atom-photon entanglement QI, quantum teleportation from photon[35,36] to matter[37] and entanglement generations for individual elementary links[38,39] have been demonstrated. However, it has been recognized that DLCZ-type QRs based on single-mode storages have very slow rates for practical use[4,40-44]. To overcome this problem, a promising way is to use multimode QMs (nodes) instead of single-mode ones to increase the elementary entanglement-generation rate[42-46]. If the quantum repeaters use the LMQIs capable of storing $N$ modes as its nodes, repeater rate will be increased by $N$ factor compared with that use single-mode LMQIs as the nodes. In recent years, temporally[47-52,59-61], spatially[46,53-56] and spectrally[45,57,58] multiplexed storages of weak coherence light or optical quantum states have been successfully demonstrated with solid-state and gas-state ensemble of atoms. With rare-earth-ion-doped (REID) crystals[60-63] (solid-state atomic ensembles), spin-wave-photon quantum correlations in more than 10 temporal modes have been demonstrated[60,61] via DLCZ approach. Based on multiple spatial-channel collection at the same time, multiplexed QI that generate entanglement between a spin-wave qubit and a photonic qubit in 6 modes have been demonstrated with a cold atomic ensemble[46]. By using a two-dimensional acoustic-optic deflector to vary spatial direction of write laser beam, L. Duan's group demonstrate a multiplexed DLCZ memory with 225 individually memory cells (modes) in a cold atomic ensemble[56]. By applying a train of write pulses in time, with each pulse coming from a different direction, to induce DLCZ Raman processes, multiplexed spin-wave-photon entanglement with 19 temporal storage modes have been achieved[59]. Also, by applying a reversible external magnetic field with a gradient to control the rephasing of spin waves, H. Riedmatten group demonstrates two time-separated spin waves in a cold atomic ensemble[64]. When comparing the temporal and spatial multimode LMQIs, one can see that the former collect and detect multi-mode quantum light field in a single optical channel, while, the later in $M$ channels. Thus, the former is easier to use. However, the temporal multimode LMQIs have a weakness, compared with the spatially-multiplexed LMQIs[65,66], that is additional noise induced by the temporal-multimode spin waves. The additional noise linearly increases with the storage mode number $M$. So it significantly decreases atom-photon quantum correlations and entanglement fidelity when the multiplexed memories have large-scale mode numbers. For example, in a temporal-multimode spin-wave-photon entanglement experiment[59], the dependence of Bell parameter $S$ as a function of the mode number m shows that $S$ linearly decrease with $m$. More precisely, when $m$=1, the measured $S$=2.65; while, when $M$=15, the measured $S$=2.35. In the construction of QRs, the improvement on entanglement fidelity is crucial and required. Many experiments[67-70] demonstrated entanglement purity from two initial entanglement sources. In several representative experiments, the entanglement-fidelity ($F$) improvements of the purified state are in a range of 5%-10%, compared with the original entanglement states[67-69]. Returning to the temporally-multiplexed DLCZ-type quantum memories, the mechanism of the additional noise generated in the multimode operations was initially pointed out by C. Simon et al.[44]. They also pointed out that the additional noise may be suppressed by using an optical cavity resonating with "write-out" photons but not resonating with "read-out" photons. Although the proposed scheme[44] can effectively reduce the additional nose, it excludes cavity-enhanced readout[27]. Following the C. Simon's scheme that use a cavity to reduce the additional noise, the group led by H. Riedmatten experimentally demonstrated the improvement of nonclassical atom-photon correlation in temporally-multiplexed of DLCZ QM[71]. So far, the improvement of entanglement fidelity of temporally-multiplexed atom-photon QI has not been reported.

Here, we propose and experimentally demonstrate an approach to suppress the additional noise. First, we generate entanglement between a temporal-multimode spin-wave QM and a Stokes photon following our previous experimental scheme[59], where, symmetric channel was set up to collect the write and read photons. The approach set up an asymmetric channel collect the write-out and read-out photons (see Fig. 1b). For making comparisons, we also set up a QI that uses symmetric channel for the photon collections. When the QIs store 14 modes, the measured Bell parameter $S$ for the QIs using the asymmetric and the symmetric photon-collection channels are 2.36±0.03 and 2.24±0.04, respectively, showing that the QI using the asymmetric channel gives rise to a 3% increase in entanglement fidelity, i.e., a 1.7-fold decrease in the additional noise, compared with the QI using the symmetric one. Contrast to the previous scheme[44], the presented temporally-multiplexed scheme allows one to improve retrieval efficiency via a cavity-enhanced atom-photon coupling[27].

**RESULTS**

**Experimental scheme of the temporal-multimode memory noise suppression.**

We now describe a scheme to suppress the additional noise in the DLCZ-like temporally-multiplexed quantum memory and then improve the fidelity of the temporal-multimode atom-photon entanglement. The schematic experimental setup is shown in Fig. 1a. The relevant atomic levels are shown in Fig. 1d), where, $|g\rangle = |5S^{1/2}, F = 1\rangle$, $|s\rangle = |5S^{1/2}, F = 2\rangle$ $|e_1\rangle = |5P^{1/2}, F' = 2\rangle$ and $|e_2\rangle = |5P^{1/2}, F' = 1\rangle$. We create the temporal-multiplexed QM that is entangled with a photon following our previous work[50]. After the atoms are released from the magneto-optical trap, we prepare the atoms into the ground state $|g\rangle$ and then start the temporal-multimode spin-wave-photon entanglement (SWPE) generation. In each generation trial, we apply a train of write pulses labeled as $W(t_i)$ $(i = 1\ to\ m)$, with each pulse coming from a different direction, to a cloud of cold atoms. The write laser is $\sigma^+$-polarized and blue detuned from the $|g\rangle \rightarrow |e_1\rangle$ transition by 20 MHz. Each write pulse, for example, the $W(t_i)$ pulse creates a spin wave in spatial mode $M(t_i)$ and simultaneously a Stokes (write-out) photon in time bin $S(t_i)$. As shown in Fig. 1a, we collected Stokes photons in two channels, one of which is asymmetric collection channel (see Fig. 1b) and another symmetric collection channel (see Fig. 1c), which are labeled as $CH_\alpha$ with $\alpha = 1, 2$. The Stokes (write-out) photons created at the $i$-th time bin and collected in the $CH_\alpha$ are denoted as $S_\alpha(t_i)$. The spin wave associated with the creations of the photon time bins $S_\alpha(t_i)$ are denoted as $M_\alpha(t_i)$, whose wave vectors are defined as $k_{M_\alpha}(t_i) = k_w(t_i) - k_{S_\alpha}$, where, $k_w(t_i)$ is wave vector of the write pulse $W(t_i)$, $k_{S_\alpha}$ are wave vector of the Stokes photon in any one of time bins in the $CH_\alpha$ channel. Furthermore, considering the polarization correlation between the Stokes photon and the internal states of the atomic spin-wave excitation[59], the atom-photon joint state, which is created in $CH_\alpha$ channel by $W(t_i)$, may be written as:

$$\rho_\alpha^{(i)} = |0\rangle_\alpha^{(i)(i)}\langle 0| + \sqrt{\chi_\alpha^{(i)}}|\Phi\rangle_\alpha^{(i)(i)}\langle\Phi|, \tag{1}$$

where, $\alpha = 1\,or\,2$ denotes $CH_1$ or $CH_2$ channel, $|0\rangle_\alpha^{(i)}$ ($\langle 0|_\alpha^{(i)}$) denotes the vacuum part of the $\alpha - th$ channel, $\chi_\alpha^{(i)}(\ll 1)$ is the probability of generating one pair of a Stokes photon and a spin wave in $i$-th mode of the $\alpha$-th channel,

$$|\Phi_{ap}\rangle_\alpha^{(i)} = \frac{1}{\sqrt{2}}\left(sin\vartheta|R\rangle_{S_\alpha}^{(i)}|\uparrow\rangle_{M_\alpha}^{(i)} + \cos\vartheta|L\rangle_{S_\alpha}^{(i)}|\downarrow\rangle_{M_\alpha}^{(i)}\right) \tag{2}$$

denotes $i$-th atom-photon entanglement state in $\alpha - th$ channel, $|R\rangle_{S_\alpha}^{(i)}$

$(|L\rangle_{S_\alpha}^{(i)})$ denotes the $\sigma^+$ ($\sigma^-$) -polarized Stokes photon in time bin $S_\alpha(t_i)$, $|\uparrow\rangle_{M_\alpha}^{(i)}$ ($|\downarrow\rangle_{M_\alpha}^{(i)}$) denotes one spin-wave(SW) excitation in the spin-wave mode $M_\alpha(t_i)$, $\cos\vartheta$ is the relevant Clebsch-Gordan coefficient with the asymmetric angle of $\vartheta \approx 0.81\frac{\pi}{4}$. Based on the Eq. (2), one can see that the spin-wave-photon entanglement (SWPE) can be created in each mode. So, our multiplexed quantum interface storing $m$ spin-wave modes may generate SWPE with a total probability

$$\chi_\alpha^{(m)} = \chi_\alpha^{(1-\text{th})} + \chi_\alpha^{(2-\text{th})}...+\chi_\alpha^{(m-\text{th})} \approx m\chi_\alpha \qquad (3)$$

where, we have assumed $\chi_\alpha^{(1-th)} \simeq \chi_\alpha^{(2-th)}... \approx \chi_\alpha^{(m-th)} \approx \chi_\alpha$ since the excitation probabilities for various $M_\alpha(t_i)$ modes are approximately equal in our presented experiment.

We use single-photon detector $D_w^\alpha$ ($D_{w_1}^\alpha$ or $D_{w_2}^\alpha$, which is shown in Fig. 1a, cf. below for details) to detect the Stokes photon. As shown in time sequence of Fig. 1e, the counts of the detectors $D_{w_1}^\alpha$ and $D_{w_2}^\alpha$ are registered only during the time bins $S_\alpha(t_1)\cdots S_\alpha(t_i)\cdots S_\alpha(t_m)$. When a Stokes photon is detected in the mode $S_\alpha(t_i)$, the storage of the spin wave in the mode $k_{M_\alpha}(t_i) = k_w(t_i) - k_{S_\alpha}(t_i)$ is heralded and the read laser pulse $R_i$ with its frequency on the $|s\rangle \to |e_2\rangle$ transition and its direction along $-k_w(t_i)$ is switched on by a feed-forward signal produced from a field-programmable gate array (FPGA), which convert the spin wave $k_{M_\alpha}(t_i)$ into an anti-Stokes photon with the wave vector $k_{AS\alpha} \approx -k_{S\alpha}$, where $k_{S\alpha}(t_1) = k_{S\alpha}(t_2) = ... = k_{S\alpha}(t_m)$, meaning that the retrieved anti-Stokes photon propagates along the opposite direction of the Stokes photon $S_\alpha(t_i)$. If two Stokes photons are detected in the modes $S_\alpha(t_l)$ and $S_\alpha(t_k)$ during one write pulse train is applied ($l, k \in m$ and $l < k$) , only the detection event in $S_\alpha(t_l)$ is registered by the FPGA. Controlled by the signal from the FPGA, a reading laser $R_l$ with direction $k_{R_l} = -k_{w_l}$ is switched on and the excitation in the $M_\alpha(t_l)$ mode is converted into an anti-Stokes photon. While, the detection event in the time bin $S_\alpha(t_k)$ is neglected. At $CH_\alpha$ channel, the Stokes (anti-Stokes) photon emitted (retrieved) from the atoms is initially collected by a fiber collimator labeled as $FC_{S\alpha}$ ($FC_{AS\alpha}$) in Fig. 1b (Fig. 1c) and then directed into an optical filter composed of several Fabry-Perot etalons. After the etalons, the $\sigma^+(\sigma^-)$-polarized Stokes (anti-Stokes) photon in $CH_1$ or $CH_2$ channel are transformed into the $H(V)$-polarized photon after passing a $\lambda/4$ plate. Then, the Stokes (anti-Stokes) photon passes through a phase compensator (PCs labeled in Fig. 1a), which can eliminate the phase shifts between the $H$- and $V$-polarized light fields resulting from the optical elements such as optical fibers and filters. Finally, the Stokes (write-out) photon or anti-Stokes (read-out) photon is guided into a polarization-beam-splitter which transmits horizontal polarization and reflects vertical polarization into single photon detectors $D_{w_1}^\alpha$ and $D_{w_2}^\alpha$ or $D_{r_1}^\alpha$ and $D_{r_2}^\alpha$, respectively. Actually, the atom-photon entanglement state $|\Phi_{ap}\rangle_\alpha^{(i)}$ will be transferred into a two-photon entangled state after the retrieval: $|\Phi_{pp}\rangle_{\alpha}^{i-\text{th}} = \frac{1}{\sqrt{2}}\left(\cos\vartheta|H\rangle_{S_{\alpha}}^{i-\text{th}}|H\rangle_{AS\alpha}^{i-\text{th}} + \sin\vartheta|V\rangle_{S_{\alpha}}^{i-\text{th}}|V\rangle_{AS\alpha}^{i-\text{th}}\right)$, where, $|H\rangle_{S\alpha}^{i-\text{th}}$

$(|V\rangle_{S\alpha}^{i-\text{th}})$ denotes an $H$ ($V$) -polarized Stokes photon in $S_{\alpha}(t_i)$ mode and $|H\rangle_{AS\alpha}^{i-\text{th}}$ $(|V\rangle_{AS\alpha}^{i-\text{th}})$ denotes a $H$ ($V$) -polarized anti-Stokes photon retrieved from the spin wave $M_{\alpha}(t_i)$. Noted that the $\sigma^{+}(\sigma^{-})$-polarization of the Stokes (anti-Stokes) photons in CH$\alpha$ channel are transformed into the $H$ ($V$) -polarization by $\lambda/4$ plate mentioned above.

As pointed out by the previous works[59], although the temporally-multiplexed QI storing $m$ spin waves promises a $m$-fold increase in the probability of generating an atom-photon entanglement state, compared with the single-mode QI, the unwanted spin waves will be also created in the temporal-multimode spin-wave memory. These unwanted spin waves will be converted into additional noise during the read-out process. We now describe the additional noise in the following. The probability of detecting a Stokes photon in one time bin at CH$\alpha$ channel is written as

$$P_{S_{\alpha}} = \chi_{\alpha}\eta_w \qquad (4)$$

where, $\eta_w$ is detection efficiency of Stokes photon at CH$\alpha$ channel. The probability of detecting an anti-Stokes photon retrieved from the individual spin-wave mode at CH$\alpha$ channel is written as[71]:

$$P_{AS_{\alpha}} = \chi_{\alpha}\gamma_{\alpha}\eta_r + N_s(1-\gamma_{\alpha})\beta_{r,\alpha}\xi_{se}\eta_r + (m-1)N_s\beta_{r,\alpha}\xi_{se}\eta_r \qquad (5)$$

where, $\gamma_{\alpha}$ is intrinsic efficiency of retrieving spin wave in the individual mode at CH$\alpha$ channel, $m$ is the spin-wave mode number, which equal to the number of the applied write pulses, $\beta_{w,\alpha}$ ($\beta_{r,\alpha}$) is the fraction of solid angle corresponding to the Stokes (anti-Stokes) photon collection mode, $N_s = \chi_\alpha / \beta_{w,\alpha}$ is the total number of created spin excitations per write pulse in the atomic ensemble[71], $\eta_r$ is detection efficiency of anti-Stokes photon at CH$_\alpha$ channel, $\xi_{se}$ is the branching ratio corresponding to the read photon transition, which is on the order of $10^{-1}$ in the presented experiment. With the $N_s = \chi_\alpha / \beta_{w,\alpha}$, we rewrite the probability $P_{AS_\alpha}$ in Eq. (5) as:

$$P_{AS_\alpha} = \chi_\alpha \gamma_\alpha \eta_r + \chi_\alpha \left(1 - \gamma_\alpha\right) \frac{\beta_{r,\alpha}}{\beta_{w,\alpha}} \xi_{se} \eta_r + \left(m-1\right) \chi_\alpha \frac{\beta_{r,\alpha}}{\beta_{w,\alpha}} \xi_{se} \eta_r \qquad (6)$$

In the above expression, the first term ($\chi_\alpha \gamma_\alpha \eta_r$) denotes the signal photon retrieved from the heralded spin wave, the second term denotes the noise due to the un-imperfect retrieval efficiency, the last term denotes the additional noise resulting from non-direction emissions of the unwanted spin waves induced by *m*-1 write pulses. The probability of detecting a coincidence between the Stokes and anti-Stokes photons in each mode at CH$_\alpha$ channel is written as:

$$P_{S,AS}(\alpha) = \chi_\alpha \gamma_\alpha \eta_w \eta_r + P_{S_\alpha} P_{AS_\alpha} \qquad (7)$$

where, $P_{S_\alpha} P_{AS_\alpha} = \chi_\alpha^2 \gamma_\alpha \eta_w \eta_r + \chi_\alpha^2 \eta_w \left(1 - \gamma_\alpha\right) \frac{\beta_{r,\alpha}}{\beta_{w,\alpha}} \xi_{se} \eta_r + \chi_\alpha^2 \eta_w \left(m-1\right) \frac{\beta_{r,\alpha}}{\beta_{w,\alpha}} \xi_{se} \eta$ denotes an accidental probability in the coincidence. In this expression, the first and second terms correspond to an accidental event in single-mode memory; the third one is the accidental event due to (*m*-1)-mode memories. The Bell parameter $S$ is given by $S = S_{max}V$, where, $S_{max} = 2\sqrt{2}$, $V$ is visibility of the interference fringe between the anti-Stokes and Stokes photons. For the current temporally-multiplexed memories, the dependence of visibility $V$ on the mode number $m$ (see Methods for details) can be written as:

$$V_\alpha(m) \approx V_\alpha(1) \frac{\gamma_\alpha}{\gamma_\alpha + 2\chi(m-1)\xi_{se}(\beta_{r,\alpha}/\beta_{w,\alpha})} \tag{8}$$

where, $V_\alpha(1)$ is the visibility for single mode case. From the above expression, one can see that the fidelity of the atom-photon entanglement decreases with the mode number $m$. In conventional experiments on atom-photon entanglement[59], the collection of the write and read photons are symmetric, i.e., $\beta_{r,\alpha}/\beta_{w,\alpha} \approx 1$. For suppressing the additional noise in the temporal-multimode spin-wave memory, one can choose asymmetric photon collection channel $CH_\alpha$ with $\beta_{r,\alpha}/\beta_{w,\alpha} < 1$ to generate the atom-photon entanglement. The asymmetric photon collection channel ($CH_1$) is shown in Fig. 1b. We also set a symmetric photon collection channel ($CH_2$) for comparing the results between the asymmetric and symmetric channels. Fig. 1b and c show the profiles of Hermite-Gaussian 00 modes in $CH_1$ and $CH_2$. As shown in Fig. 1b, the light beam emitting from a fiber collimator (PAF-X-15-B, Thorlabs), which is labeled as $FC_{S1}$, placed at the position $x_1$ has a beam diameter of 2.6 mm, we place a 2.4 m focal length lens after the collimator. Thus, the light beam at the position $x_2$ ($x_2 - x_1 = -2.45\ m$) has a beam diameter of 1.2 mm. The center of the atoms is at the position of $x_0$ ($x_0 - x_2 = 1.25\ m$) and has a beam diameter of 1.3 mm. The light beam at $x_2$ is coupled into a fiber collimator (PAF-X-7-B, Thorlabs), which is labeled as FC$_{AS1}$. Since the collimators at $x_1$ and $x_2$ positions are used to collected the write and read photons, the ratio of the solid angles $\beta_{w,\alpha=1}\big/\beta_{r,\alpha=1} \approx \frac{A_w}{A_r} = \left(\frac{2.6}{1.2}\right)^2 \approx 4.69$, where, $A_w$ and $A_r$ are the apertures of the fiber collimator at $x_1$ and $x_2$ positions in CH$_1$. Fig. 1c shows mode profile of the symmetric channel (CH$_2$). The write and read photons are collected by the same fiber collimators (PAF-X-7-B, Thorlabs), which are labeled as FC$_{S2}$ and FC$_{AS2}$, respectively. According to the beam diameters as shown in Fig. 1c, we have $\beta_{w,\alpha=2}\big/\beta_{r,\alpha=2} = 1$.

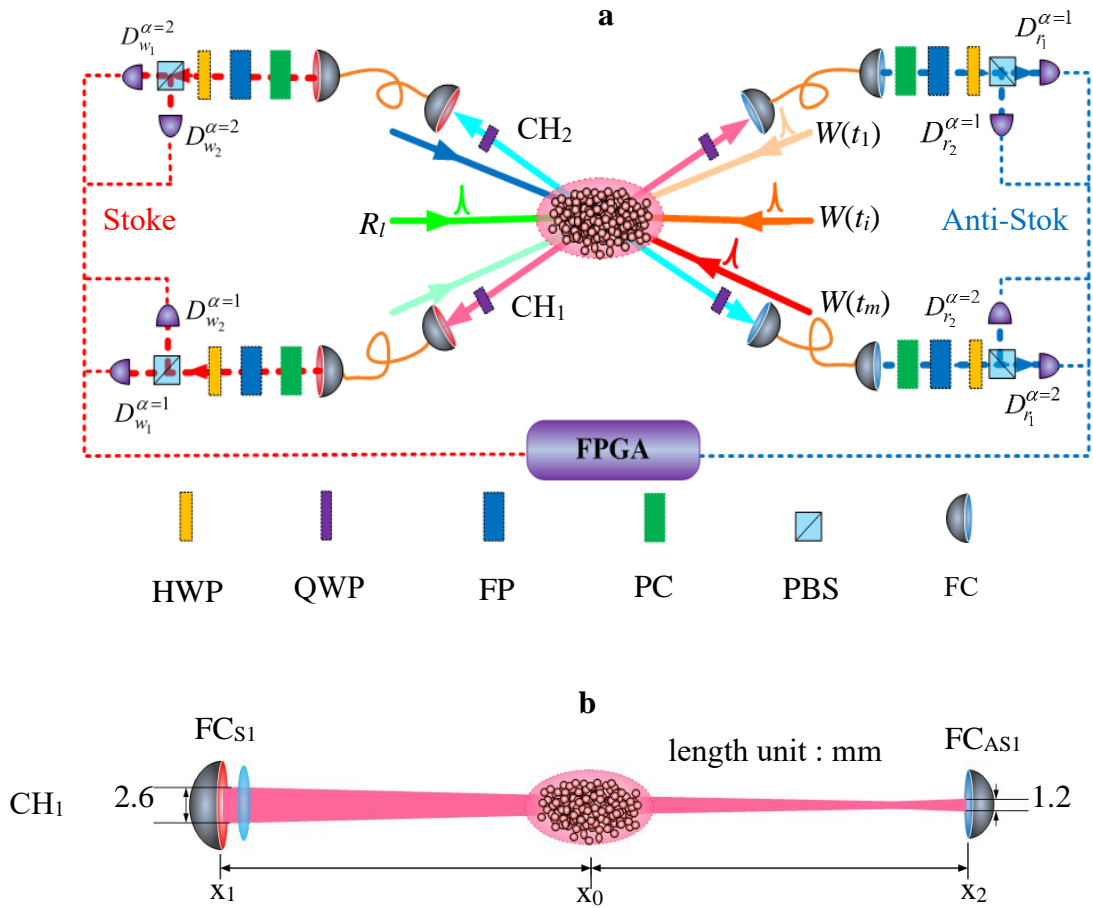

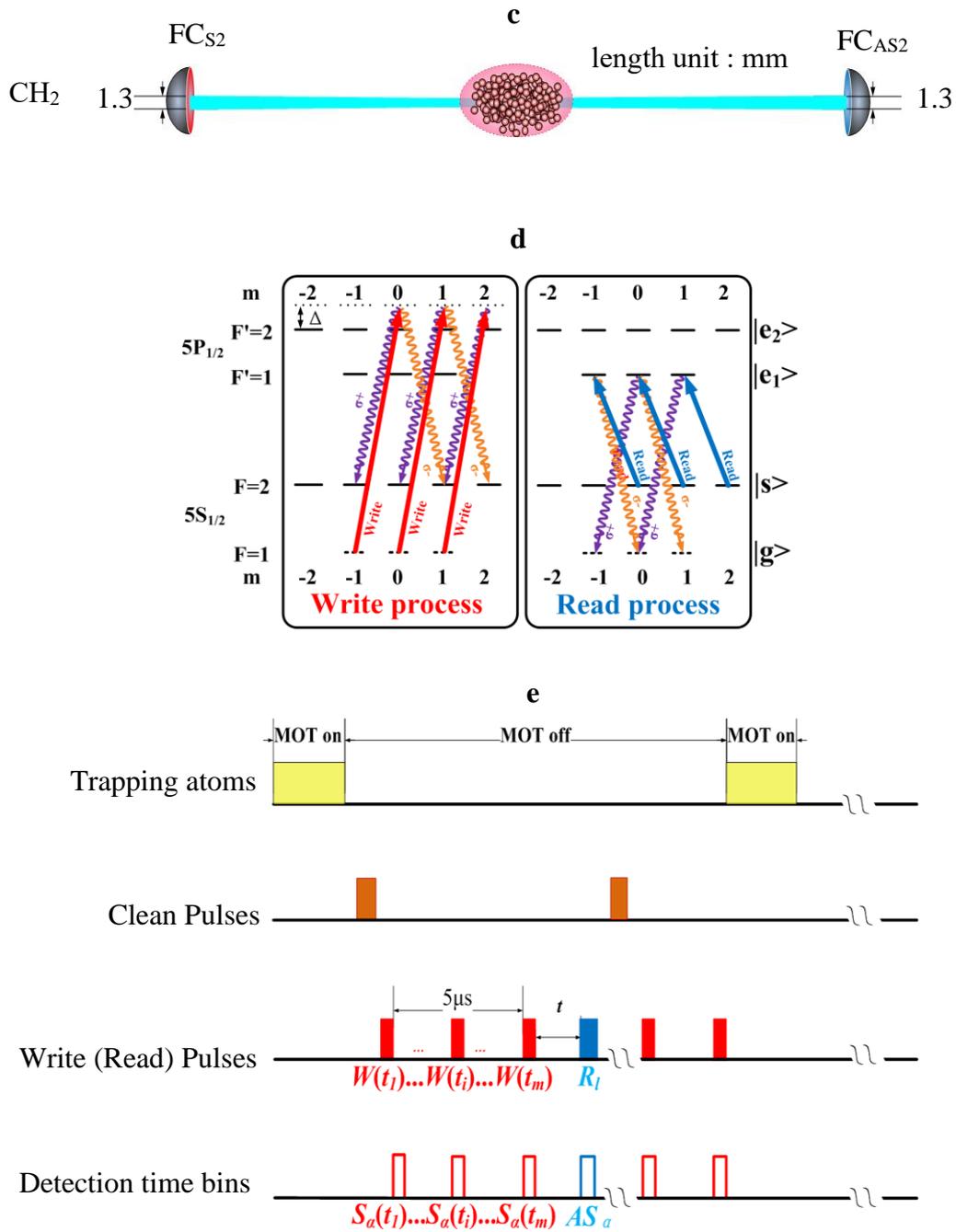

**Fig. 1 Overview of the experiment. a** Experimental setup . A train of write pulses labeled as $W(t_i)$

($i$ = 1 to m), with each pulse coming from a different direction, is applied onto a cloud of cold

atoms, with $m$ being up to 14 (for simplicity, we only plot three-directions). The asymmetric (CH$_1$)

and symmetric (CH$_2$) channels are set up for collecting Stokes and Anti-Stokes photons. FPGA:

programmable gate array; PBS: polarization-beam splitter; FC: fiber collimators; PC: phase compensation; HWP: half-wave plate; QWP: quarter-wave plate; D: single photon detector; ASα: Anti-Stokes photons; FP: Fabry-Perot filters. FC$_{S1,2}$, FC$_{AS1}$: fiber collimators; **b** asymmetric photon collection channel. **c** symmetric photon collection channel; **d** Relevant atomic levels. e Time sequence of the experimental trials.

## Experimental demonstration

As discussed in the above, by applying a train containing m write laser pulses, with each pulse coming from a different direction, to a cloud of cold atoms, we generate atom-photon (photon-photon) entanglement in m temporal modes.

The red circle and black square dots in Fig. 2 are the measured generation probabilities $\chi_{\alpha=1}^{(m)}$ and $\chi_{\alpha=2}^{(m)}$ as a function of the mode number m for 100 $\mu W$ write power. The measured results show that probabilities $\chi_{\alpha=1}^{(m)}$ and $\chi_{\alpha=2}^{(m)}$ linearly increases with m, i.e., that the QI storing m spin-wave modes may increase the probability of generating atom-photon entanglement by m factor. On the other hand, the atom-photon entanglement excitation probability ($\chi_{\alpha=1}^{(m)}$) in CH$_1$ is significantly larger than that ($\chi_{\alpha=2}^{(m)}$) in CH$_2$. The reason for this is that the Stokes collection angle in CH$_1$ channel is larger than that in CH$_2$.

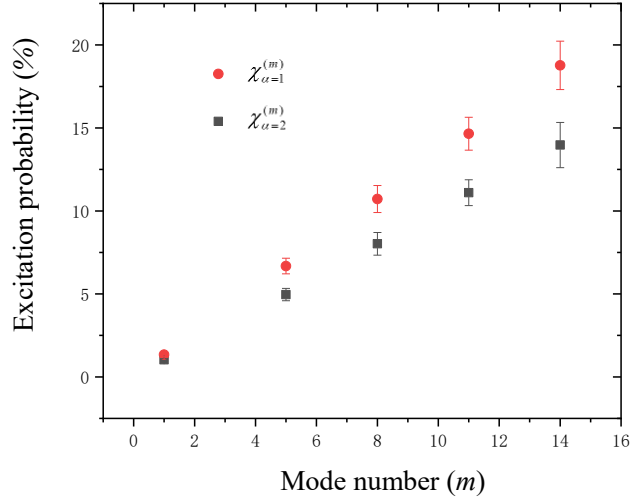

**Fig. 2 Measured excitation probabilities** $\chi_{\alpha=1}^{(m)}$ **(red circle dots) and** $\chi_{\alpha=2}^{(m)}$ **(black square dots) as a function of mode number $m$.** Error bars in the experimental data represent 1 standard deviation, which is estimated from the Poissonian detection statistics.

We also measured the retrieval efficiencies of spin waves $M(t_i)$ in CH$_1$ and CH$_2$ channel and find they are basically symmetric. The average retrieval efficiency in CH$_1$ channel is 15.8% and that in CH$_2$ is 16.7%.

The quality of the atom-photon (photon-photon) entanglement created from the temporally-multiplexed QI storing $m$ spin-wave modes can be characterized by the Bell parameter

$$S_{\alpha}^{(m)} = \left| E^{(m)}(\theta_S, \theta_{AS}) - E^{(m)}(\theta_S, \theta_{AS}^{'}) + E^{(m)}(\theta_S^{'}, \theta_{AS}) + E^{(m)}(\theta_S^{'}, \theta_{AS}^{'}) \right| < 2$$ , with the correlation function $E_{\alpha}^{(m)}(\theta_S, \theta_{AS})$ given by

$$\frac{C_{D_{w_1}^\alpha, D_{r_1}^\alpha}^{(m)}(\theta_S, \theta_{AS}) + C_{D_{w_2}^\alpha, D_{r_2}^\alpha}^{(m)}(\theta_S, \theta_{AS}) - C_{D_{w_1}^\alpha, D_{r_2}^\alpha}^{(m)}(\theta_S, \theta_{AS}) - C_{D_{w_2}^\alpha, D_{r_1}^\alpha}^{(m)}(\theta_S, \theta_{AS})}{C_{D_{w_1}^\alpha, D_{r_1}^\alpha}^{(m)}(\theta_S, \theta_{AS}) + C_{D_{w_2}^\alpha, D_{r_2}^\alpha}^{(m)}(\theta_S, \theta_{AS}) + C_{D_{w_1}^\alpha, D_{r_2}^\alpha}^{(m)}(\theta_S, \theta_{AS}) + C_{D_{w_2}^\alpha, D_{r_1}^\alpha}^{(m)}(\theta_S, \theta_{AS})} ,$$

where, for example, $C_{D_{w_1}^\alpha, D_{r_1}^\alpha}^{(m)}(\theta_S, \theta_{AS}) = \sum_{t=1}^{m} C_{D_{w_1}^\alpha, D_{r_1}^\alpha}^{(i\text{-th})}(\theta_S, \theta_{AS})$ , $C_{D_{w_1}^\alpha, D_{r_1}^\alpha}^{(i\text{-th})}(\theta_S, \theta_{AS})$

denotes the coincidence counts between detecting a Stokes photon in the time-bin $S_{\alpha(t_i)}$ by $D_{w_i}^{\alpha}$ and detecting an retrieved anti-Stokes photon from $M_{\alpha}(t_i)$ by $D_{r_i}^{\alpha}$ for the polarization angles $\theta_S$ and $\theta_{AS}$ at CH$\alpha$ channel. In the measurement, the canonical settings are chosen to be $\theta_s = 0°$, $\theta_s^{'} = 45°$, $\theta_{AS} = 22.5°$ and $\theta_{AS}^{'} = 67.5°$. The red circle and blue diamond dots in Fig. 3 depict the measured $S_{\alpha=1}^{(m)}$ and $S_{\alpha=2}^{(m)}$ data when the memory stores $m$ modes. It shows that the measured Bell parameters $S_{\alpha=1}^{(m=14)}$ and $S_{\alpha=2}^{(m=14)}$ are $2.36 \pm 0.03$ and $2.24 \pm 0.04$, showing that the asymmetrical configuration increases Bell parameter $S$ by 0.12 compared with symmetrical one. The red dash line and blue solid line are the fits to the results of CH$_1$ and CH$_2$ based on Eq. (8) with $\beta_{w,\alpha=1}/\beta_{r,\alpha=1}=1.7$ and $\beta_{w,\alpha=2}/\beta_{r,\alpha=2}=1$. One can see that the ratio value $\beta_{w,\alpha=1}/\beta_{r,\alpha=1}\approx1.7$ using in the experimental fitting is not in agreement with the theoretically evaluated value of $\beta_{w,\alpha=1}/\beta_{r,\alpha=1}\approx4.69$. Currently, we attribute this disagreement to the light-field mode difference between our present experiment and the theoretical proposal. More precisely, the light field in the form is Hermite-Gauss 00 mode and that in the later is spherical wave. In the coincidence measurements, generation probabilities $\chi_{\alpha=1}$ and $\chi_{\alpha=2}$ are both chosen to be ~1%, which are obtained by setting the write power to be 75 $\mu W$ and 100 $\mu W$, respectively.

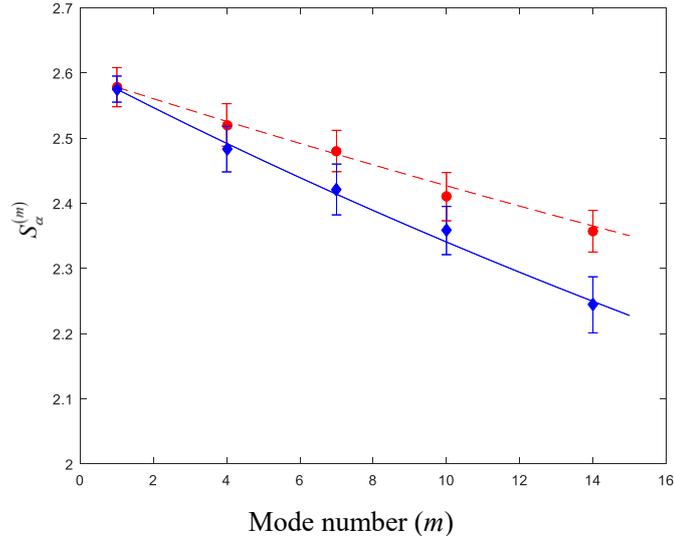

**Fig. 3 Measurements of the Bell parameter** $S_\alpha^{(m)}$ **as a function of mode number** *m*. With the excitation probabilities $\chi_{\alpha=1} \approx \chi_{\alpha=2} \approx 1\%$. Error bars in the experimental data represent 1 standard deviation, which is estimated from the Poissonian detection statistics. The red dash (blue solid) line is the fitting to the measured data $S_\alpha^{(m)}$ based on Eq. (8) with the parameters of branching ratio $\xi_{se} = 0.093$, retrieval efficiencies $\gamma_{\alpha=1} = 15.8\%$, $\gamma_{\alpha=2} = 16.7\%$, ratios of the solid angles $\beta_{w,\alpha=1}/\beta_{r,\alpha=1} = 1.7$, $\beta_{w,\alpha=2}/\beta_{r,\alpha=2} = 1$, visibilities for single mode $V_{0_{\alpha=1}}^{'} \approx V_{0_{\alpha=2}}^{'} \approx 0.91$ (see Methods for details).

To investigate the multimode storage ability of the ensemble, we measure the decay of the Bell parameter $S_{\alpha=1}^{(m=14)}$ and $S_{\alpha=2}^{(m=14)}$ with the storage time *t* for the generation probabilities $\chi_{\alpha=1} \approx \chi_{\alpha=2} \approx 1\%$, respectively. The red circle (black square) dots in Fig. 4 depict the measured $S_\alpha^{(m=14)}$ as a function of storage time *t*. In CH$_1$ channel, the measured Bell parameter $S_{\alpha=1}^{(m=14)} = 2.12 \pm 0.04$ for storage time *t* = 25 $\mu s$, violating the Bell inequality by 3 standard deviations. In CH$_2$ channel, the measured $S_{\alpha=2}^{(m=14)} = 2.06 \pm 0.03$ for storage time *t* = 20 μs, violating the Bell inequality by 2 standard deviations.

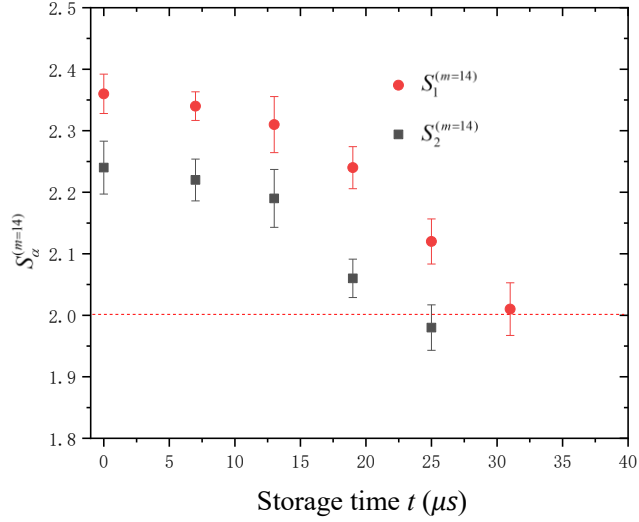

**Fig. 4 Bell parameter** $S_\alpha^{(m=14)}$ **as a function of storage time** *t*. With the excitation probability $\chi_{\alpha=1} \approx \chi_{\alpha=2} \approx 1\%$. The red circle (black square) dots depict the measured $S_\alpha^{(m=14)}$ for various storage times *t* in CH$_1$(CH$_2$) channel. Error bars in the experimental data represent 1 standard deviation, which is estimated from the Poissonian detection statistics.

## Discussion

We experimentally demonstrate a scheme that can improve the entanglement quality of temporal multimode atom-photon QI by using an asymmetric photon-collection channel instead of symmetric one. When the atom-photon entanglement QIs store 14 temporal modes, the measured Bell parameter *S* are $2.24\pm0.04$ and $2.36\pm0.03$ for QI using the symmetrical and asymmetrical channels, respectively, showing that the QI using the asymmetric channel gives rise to a 3% increase in entanglement fidelity, i.e., a 1.7-fold decrease in the additional noise, compared with the QI using the symmetric one. Based on our theoretical expectation (see Methods for details), the storage temporal mode number in multiplexed QI using the asymmetrical channel promising to violate the Bell inequality is 42, which is far beyond the value of 26 modes for the multiplexed QI using symmetrical one. The presented temporal multimode DLCZ-like memories can be combined with the spatial multiplexing scheme[41]and then promise one to achieve large-scale multiplexed QIs. Considering a multiplexed atom-photon entanglement source that stores 14 temporal and 15 spatial SW qubits, the total number of memory qubits will reach $N_m = 210$. Moreover, the presented temporally-multiplexed scheme allows one to improve retrieval efficiency via a cavity-enhanced atom-photon coupling. The short storage lifetime (~ 25 $\mu s$) can be extended by trapping atoms in an optical lattice[15,17] and selecting two magnetic-field-insensitive spin waves to store memory qubits[72,73]. To minimize transmission losses in fibers, one can convert the Stokes photons (795 nm) into photons in the telecommunications band[74-76]. The presented work paves a road to achieve high-performance LMQI and will benefit QR-based long-distance quantum communications.

## Methods

The visibility of the atom-photon entanglement QI storing *m* modes in CH$\alpha$ can be written as $V_\alpha(m) = V_{0_\alpha}^{'} \frac{P_{S,AS}(\alpha) - P_{S_\alpha} P_{AS_\alpha}}{P_{S,AS}(\alpha) + P_{S_\alpha} P_{AS_\alpha}}$ ,where, $V_{0_\alpha}^{'}$ is the single-mode visibility, whose value is less than 1, which is reasonable for the imperfect phase compensation of the optical elements and the asymmetry angle $\vartheta$ described in Eq. (2). According to the expressions of $P_{S,AS}(\alpha)$ , $P_{S_\alpha}$ and $P_{AS_\alpha}$ , we rewrite the visibility $V_\alpha(m)$ as:

$$V_\alpha(m) \approx \frac{V_{0_\alpha}^{'}}{1 + 2\left(\chi_\alpha + \frac{\chi_\alpha}{\gamma_\alpha}(1-\gamma_\alpha)\xi_{se}(\beta_{r,\alpha}/\beta_{w,\alpha})\right)} \frac{1}{1 + \frac{2(m-1)\chi_\alpha \xi_{se}(\beta_{r,\alpha}/\beta_{w,\alpha})}{\gamma_\alpha} \Big/ \left(1 + 2\left(\chi_\alpha + \frac{\chi_\alpha}{\gamma_\alpha}(1-\gamma_\alpha)\xi_{se}(\beta_{r,\alpha}/\beta_{w,\alpha})\right)\right)}$$

,where $V_\alpha(m=1) = \frac{V_{0\alpha}^{'}}{1 + 2\left(\chi_\alpha + \left(\chi_\alpha(1-\gamma_\alpha)\xi_{se}(\beta_{r,\alpha}/\beta_{w,\alpha})\right)/\gamma_\alpha\right)}$ is the visibility of a single mode, and then the visibility $V_\alpha(m)$ can be rewritten as

$$V_\alpha(m) = \frac{V_\alpha(1)}{1 + \frac{2\chi_\alpha(m-1)\xi_{se}(\beta_{r,\alpha}/\beta_{w,\alpha})}{\gamma_\alpha} \Big/ \left(1 + 2\left(\chi_\alpha + \frac{\chi_\alpha}{\gamma_\alpha}(1-\gamma_\alpha)\xi_{se}(\beta_{r,\alpha}/\beta_{w,\alpha})\right)\right)}.$$

Since $\chi_\alpha + \frac{\chi_\alpha}{\gamma_\alpha}(1-\gamma_\alpha)\xi_{se}(\beta_{r,\alpha}/\beta_{w,\alpha}) \approx 0$ , we have $V_\alpha(m) = \frac{V_\alpha(1)}{1 + 2\chi_\alpha(m-1)\xi_{se}(\beta_{r,\alpha}/\beta_{w,\alpha})/\gamma_\alpha}$ .

Theoretically calculated Bell parameters $S_\alpha^{(m)}$ as a function of $m$ according to Eq. (8) with the same parameters as those in Fig. 3.

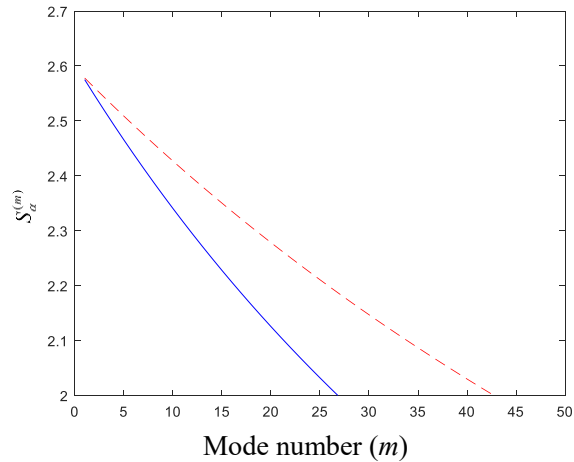

**Fig. 5 Theoretically calculated the Bell parameters**. $S_1^{(m)}$ (red dash curve) and $S_2^{(m)}$ (blue solid

curve) as a function $m$ based on Eq. (8), where the parameters used in Fig. 5 are the same with that in Fig. 3, i.e., excitation probability $\chi_{\alpha=1} \approx \chi_{\alpha=2} \approx 1\%$, branching ratio $\xi_{se} = 0.093$, retrieval efficiencies $\gamma_{\alpha=1} = 15.8\%$ and $\gamma_{\alpha=2} = 16.7\%$, the ratios of the solid angles $\beta_{w,\alpha=1} / \beta_{r,\alpha=1} = 1.7$ and $\beta_{w,\alpha=2} / \beta_{r,\alpha=2} = 1$, the visibilities for single mode $V'_{0_{\alpha=1}} \approx V'_{0_{\alpha=2}} \approx 0.91$.

# DATA AVAILABILITY

All the data and calculations that support the findings of this study are available from the corresponding author upon reasonable request.

# CODE AVAILABILITY

The code used to generate data will be made available to the interested reader upon reasonable request.

## ACKNOWLEDGEMENTS

We acknowledge funding support from Key Project of the Ministry of Science and Technology of China (Grant No. 2016YFA0301402); The National Natural Science Foundation of China (Grants: No. 11475109, No. 11974228). Fund for Shanxi "1331 Project" Key Subjects Construction.

## AUTHOR CONTRIBUTIONS

H. W. conceived the research. H. W., Y.L. and Y.-F. W. designed the experiment. Y.L., Y.-F.W. , M.-J. W. setup the experiment with assistances from all other authors. Y.L. and Y.-F. W. took the data. Y.L. analyzed the data. H. W. wrote the paper.

## COMPETING INTERESTS

The authors declare no competing interests.